\documentclass[%
reprint,
superscriptaddress,
showpacs,preprintnumbers,
amsmath,amssymb,
aps,pre, 
nofootinbib,
longbibliography, 
floatfix,
]{revtex4-1}

\usepackage{enumitem}  
\usepackage{graphicx}
\usepackage{epstopdf}
\usepackage{dcolumn}
\usepackage{bm}
\usepackage{color,soul}
\usepackage[none]{hyphenat}

\newcommand{\ie}{\textit{i.e.,}\ }
\newcommand{\eg}{\textit{e.g.}\ }

\begin{document}

\title{Optical phase conjugation in backward Raman amplification}

\author{Qing Jia}
\affiliation{Department of Astrophysical Sciences, Princeton University,  Princeton, New Jersey 08544, USA}  
\affiliation{Now working at the Department of Engineering and Applied Physics, University of Science and Technology of China, Hefei, 230026, China}
\author{Kenan Qu}
\affiliation{Department of Astrophysical Sciences, Princeton University,  Princeton, New Jersey 08544, USA}  
\author{Nathaniel J. Fisch}
\affiliation{Department of Astrophysical Sciences, Princeton University,  Princeton, New Jersey 08544, USA}



\begin{abstract}
Compression of an intense laser pulse using backward Raman amplification (BRA) in plasma, followed by vacuum focusing to a small spot size, can produce unprecedented ultrarelativistic laser intensities. The plasma density inhomogeneity during BRA, however, causes laser phase and amplitude distortions, limiting the pulse focusability. To solve the issue of distortion, we investigate the use of optical phase conjugation as the seed pulse for BRA. We show that the phase conjugated laser pulses can retain focusability in the nonlinear pump-depletion regime of BRA, but not so easily in the linear amplification regime.  This somewhat counter-intuitive result is because the nonlinear pump-depletion regime features a shorter amplification distance, and hence less phase distortion due to wave-wave interaction, than the linear amplification regime.
\end{abstract}


\maketitle
Generating strong laser pulses and delivering them precisely to certain target region represent two major endeavors in laser technology development. They are important in many applications, such as laser-based particle accelerators, inertial confinement fusion, laser surgery, and laser-based weapons. Advances of these two aspects of laser technologies are often dependent on each other. Intense laser pulses induce strong refraction during propagation, which, on the other hand, limits the amplification of the laser pulses themselves. State-of-the-art laser intensities are currently obtained by splitting the pulse into multiple components for amplification before compressing/recombining them either in the frequency domain, like chirped pulse amplification, or in the space domain. For the process of pulse compression/recombination~\cite{OC-Mourou2012}, it was proposed to use backward Raman amplification (BRA)~\cite{PRL-Shvets1998, PRL-Malkin1999, POP-Malkin2000,PRL-Malkin2000} in plasma rather than using solid-state optical component to avoid the thermal damage issue. BRA eliminates the major hurdle in ultrahigh peak power laser pulse compression, holding the promise, in principle,  of exawatt to zettawatt pulses~\cite{OC-Mourou2012}.

In plasma BRA, an active plasma wave mediates the laser energy transfer from a long pump pulse to a counterpropagating short seed pulse at a lower frequency. Owning to the large laser-plasma interaction rate, an initial weak seed laser pulse can gain an $e$-fold intensity increase in a few laser cycles in the linear amplification stage. 
The fast-growing seed pulse quickly depletes the pump pulse and captures the pump energy in the short seed pulse. 
In the nonlinear pump depletion stage, the seed pulse amplitude grows linearly with time while its duration decreases inversely with time. 
It is this nonlinear stage that provides the vital compression to the final ultrashort pulse. 
For $1 \mu$m-wavelength radiation, ultrafast compression of BRA can theoretically achieve nearly relativistic intensities ($10^{17}\rm W/cm^2$), which is five orders of magnitude higher than the output intensity of $\rm 10^{12} W/cm^2$ from a typical CPA compressor.

The relativistic intensity threshold in plasma can be overcome by transversely  focusing the pulse to higher intensity which is achieved outside the plasma. 
Thus, a properly shaped laser pulse, if remaining well-focused during amplification, can deliver an unprecedented ultrarelativistic intensity at the focal point after exiting the plasma~\cite{PRL-Malkin1999,POP-Malkin2000}. 
Two-dimensional numerical simulations find that the BRA is robust to a broad range of pump and seed perturbations in homogeneous plasmas~\cite{POP-Fraiman2002}. 
Even a pump laser with finite coherence could be efficiently compressed into a short pulse in the nonlinear amplification stage~\cite{2017popEdwards}. 
In BRA experiments, one of the major restrictions is often related to ensuring the seed pulse quality when it interacts with the pump laser. 
A pre-focused seed pulse may deteriorate when propagating through random plasma inhomogeneities~\cite{POP-Solodov2003, Palastro2015}. 
The reduction of seed peak intensity delays the onset of nonlinear pump depletion and costs in energy transfer efficiency. 
The scattering might also create precursors which cause unwanted premature pump depletion~\cite{PRL-Tsid2002}. 
Unfortunately, the plasma density fluctuates randomly in space and its effect is too complicated to somehow be mitigated by adjusting the laser phasing.  

\begin{figure}[h]
\centering
\includegraphics[width=0.65\linewidth]{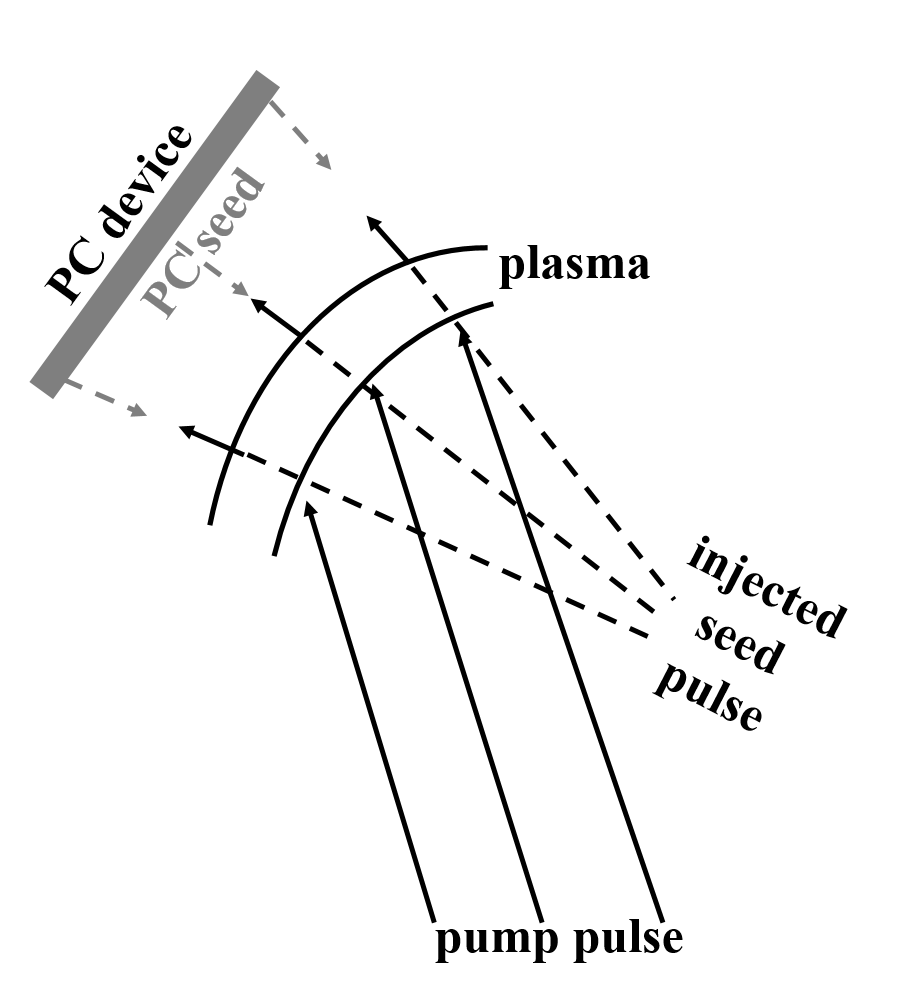}
\caption{Schematics of using optical phase conjugation of a pre-focused seed pulse in backward Raman amplification to counteract the distortion due to plasma density fluctuation.} 
\label{fig:scheme}
\end{figure}

A phase conjugation (PC) wave can compensate the phase distortion in the same random plasma by reversing the time-symmetry of pulse propagation~\cite{1977Hellwarth,1977Yariv,1979olSteel,1989prlKitagawa,1991Federici,1992Joshi,2007preLee,2012olDing}. Its implementation in seeding plasma Raman amplification could possibly avoid the vulnerability of pulse scattering by plasma density inhomogeneity. Consider the schematics in Fig.~\ref{fig:scheme}: A focused laser pulse at frequency $\omega_b$ is sent through a random plasma and reflected by a PC mirror to create the seed pulse. The PC seed pulse then propagates against a pump pulse at frequency $\omega_a$ in the same plasma. Without seed-pump interaction, the seed pulse can focus at the original focal point despite plasma inhomogeneity.  
With a proper pump, the seed pulse gets amplified through the BRA process. The condition for perfect phase correction and pulse intensity restoration using phase conjugation is that the medium is stationary and lossless. While the plasma density dynamics could be negligible during the passage of the seed pulse, it is not known whether the amplification process changes the focusability of the PC seed pulse.

To analyze the evolution of the amplified seed, we denote the PC seed pulse as $\mathbf{E}_b = \mathbf{u}_b(\mathbf{r}) e^{i\phi_b(\mathbf{r})} e^{-i(k_bx+\omega_b t)}$ with a complex amplitude $\mathbf{u}_b$ and a wavevector $k_b$. Here, $\phi_b(\mathbf{r})$ is a fluctuating phase which would gradually decrease and be perfectly compensated at the plasma boundary. 
When the PC seed interacts with a counterpropagating pump pulse $\mathbf{E}_a = \mathbf{u}_a(\mathbf{r}) e^{i\phi_a(\mathbf{r})} e^{i(k_ax-\omega_b t)}$, the ponderomotive potential of the pump-seed beating induces a plasma wave $\mathbf{E}_f = \mathbf{u}_f(\mathbf{r}) e^{i\phi_f(\mathbf{r})} e^{i(k_fx-\omega_p t)}$ if the $\omega_a-\omega_b = \omega_p$. Here, $\mathbf{u}_{a,f}$,  $k_{a,f}$, and $\phi_{a,f}(\mathbf{r})$ denote the complex amplitude, wavevector, and the random phase of the pump laser/plasma wave, respectively. The BRA process can be described through the the simplified coupled wave equations~\cite{PRL-Malkin1999,POP-Malkin2000}, 
\begin{equation} \label{11}
(\partial_t - c\partial_x ) a = -V bf, \quad
(\partial_t + c\partial_x ) b = V af^*, \quad
\partial_t f = V ab^*, 
\end{equation}
where $a$ and $b$ [ $=eu_{a,b} e^{i\phi_{a,b}(\mathbf{r})} \allowbreak/(m_ec^2\omega_{a,b})$] denote the complex envelopes of the pump and seed pulses, $f=eu_f e^{i\phi_f(\mathbf{r})} \allowbreak/(2m_ec\omega_p)$ denotes the complex envelope of the plasma wave, $V \approx \sqrt{\omega_a\omega_p}/2$ is the three-wave interaction rate, $e$ is the natural charge, $m_e$ the mass of an electron, and $c$ is the speed of light. Note that Eqs.~(\ref{11}) do not include transverse Laplacian terms because we neglect the change of interaction rate due to a small transverse phase mismatch. We assume that the transverse phase of the generated wave is solely determined by the existing waves.

\begin{figure}[bh]
	\centering
	\includegraphics[width=\linewidth]{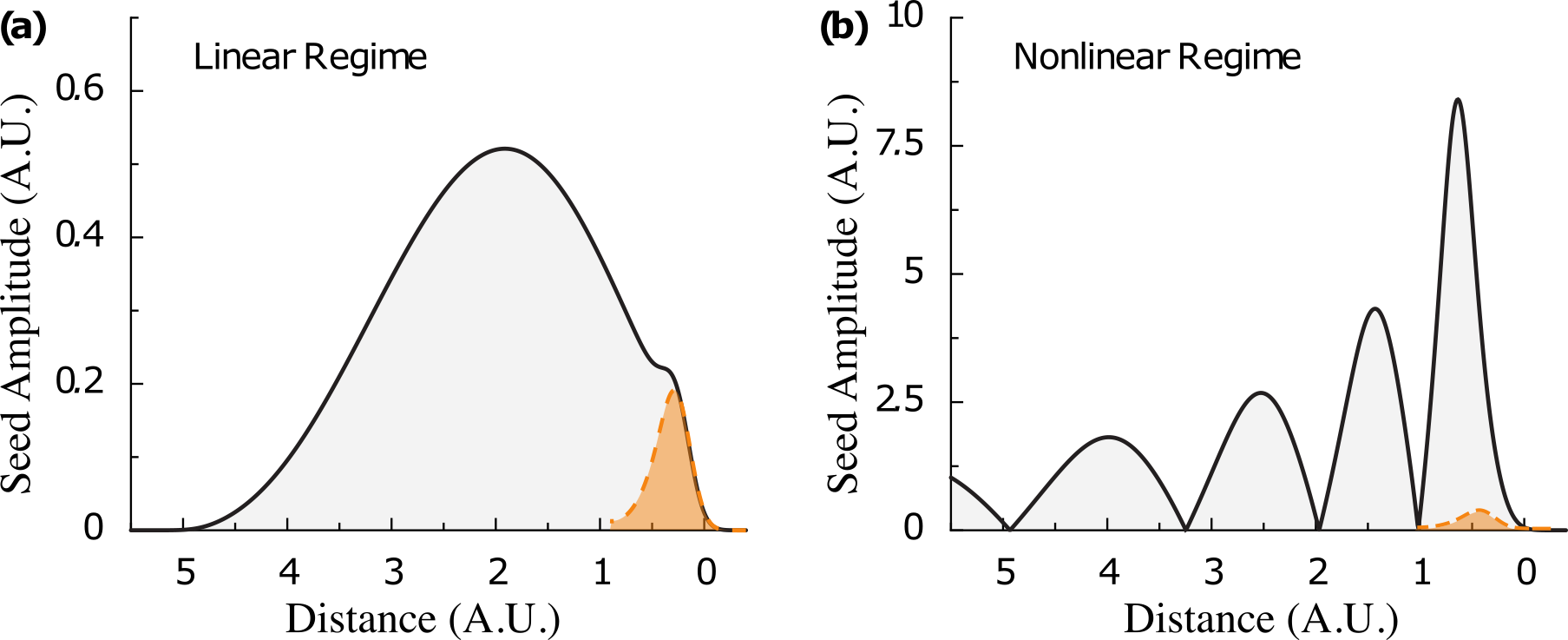}
	\caption{Comparisons of the seed pulse (orange dashed curves) and the amplified pulse (black solid curves) in the linear stage (a) and the nonlinear stage (b) of BRA, respectively.  }
	\label{fig:principle}
\end{figure}

During BRA, the generated plasma wave further interacts with the pump and causes energy transfer to the seed pulse. The seed amplitude $|\mathbf{u}_b(\mathbf{r})|$ is certainly changed during BRA. But what governs the pulse focusability is the dynamics of the laser phase $\phi_b(\mathbf{r})$. For convenience, we separately describe the seed pulse and the amplified pulse (the so-called probe pulse) as $b_0$ and $b_1$, although they have the same frequency and wavevector, \ie $b=b_0+b_1$. 
Since the plasma wave $f$ is created by the combination of the laser pulses $a$ and $b$, it has the conjugate phase of the seed, $f\sim b_0^*$. The amplified pulse $b_1$, when propagating through the plasma wave, integrates over the plasma waves with all phases. Exactly, the final amplified pulse in the linear pump-nondepletion regime ($|a|\equiv$ Const.) is~\cite{2017prlQu}
\begin{align} 
b_1(t,x) &=  \int_0^x G(t, x-x') b_0(x') dx', \label{15} \\
G(t,x) &= \sqrt{\frac{|a|^2V^2}{c^2} \frac{x}{ct-x}}\, I_1\left(\sqrt{\frac{2|a|^2V^2}{c^2} x(ct-x)} \right) , \label{16}
\end{align}
where $I_1$ is the first-order modified Bessel function.  
Equation~(\ref{15}) shows that the output amplified seed pulse, at the output $x_f=ct_f$, is the convolution of Green's function $G$ and the seed pulse $b_0$. Each part of $b_1$ after the seed pulse $b_0$ is a convolution of the whole duration of the seed pulse $b_0$, which has a random phase $e^{i\phi_b(\mathbf{r}, x)}$ at different $x$'s. Since $G$ maximizes at $x=x_f/2$, the peak of the amplified pulse $b_1$ lags behind the seed pulse, as illustrated in Fig.~\ref{fig:principle}(a). The phase of $b_1$ at its peak is different from $\phi_b(\mathbf{r})$ of the seed, and hence the amplified pulse does not retain the focusability of the PC seed in the linear stage of BRA. 

The amplification process becomes different in the nonlinear pump depletion stage which happens when the seed becomes sufficiently strong. In this stage, the strong seed quickly depletes the pump within a short interaction distance. The peak of the Green's function $G$ shifts from $x=x_f/2$ to $x\sim x_f$. As illustrated in Fig.~\ref{fig:principle}(b), the amplified pulse gains preferentially more pump energy only if it closely follows the seed pulse. The shadowing effects of the rear layer amplification due to pump depletion, controversially, benefits the pulse focusability by eliminating the unwanted convolution process. In the nonlinear regime, the transverse phase front is carried over from the PC seed pulse to the plasma wave and then immediately to the amplified pulse. The local phase of $b_1(x)$ at the leading pulse spike closely resembles the phase $\phi_b(\mathbf{r})$. Hence, the focusability of the PC seed retains in the nonlinear regime. Importantly, the pump phase fluctuation due to the plasma inhomogeneity does not affect the focusability because only the pump intensity $|a|^2$ appears in the interaction, as seen in Eqs.~(\ref{15}) and (\ref{16}).

Preparing an intense seed for reaching the nonlinear stage of BRA requires tight focusing of the laser pulse. Such a seed pulse contains a broad spectrum of wavenumbers, and each wavenumber component, after propagation in an inhomogeneous plasma, could accumulate a different phase $\phi_b(\mathbf{k},\mathbf{r})$, \ie $\displaystyle \mathbf{E}_b = \frac{1}{\sqrt{2\pi}} \int \mathbf{u}_b(\mathbf{k},\mathbf{r}) \allowbreak e^{i\phi_b(\mathbf{k},\mathbf{r})} \allowbreak e^{-i(\mathbf{k}_b\cdot\mathbf{r}+\omega_b t)} d^3k_b$. In the regions where different wavenumber components have similar phases, they constructively interfere and create local amplitude peaks; Otherwise, they destructively interfere and create local amplitude troughs. The local intensity peaks feature higher seed intensities with shorter spike duration similar to those in a multifrequency Raman amplifier~\cite{Ido_pre2018}. Since the nonlinear growth rate depends on the seed intensity, the onset of local peaks benefits rapidly reaching the nonlinear stage of amplification.

The fluid model Eqs.~(\ref{11}) assumes quasi-frequency matching and phase matching. The matching conditions, in principle, do not hold in an inhomogeneous plasma due to fluctuation of the local plasma frequency and scattering of the laser pulses. The fluctuation of plasma frequency could be compensated by the broad spectra of the local intensity peaks. The mismatch of the pulse wavefronts can, to a certain degree, also be mitigated with a PC seed pulse: Since the propagation of the PC seed pulse is a time reversal of a counter-propagating focused pulse, the wavefront of the PC seed pulse at any certain location is similar to that of the counter-propagating pump pulse when neglecting their frequency detuning and different Rayleigh lengths. However, these analysis must be checked more rigorously. In the following, we show kinetic particle-in-cell (PIC) simulations to demonstrate the advantage of PC seeds compared to other types of seeds in BRA. 

\begin{figure}[h]
\centering
\includegraphics[width=\linewidth]{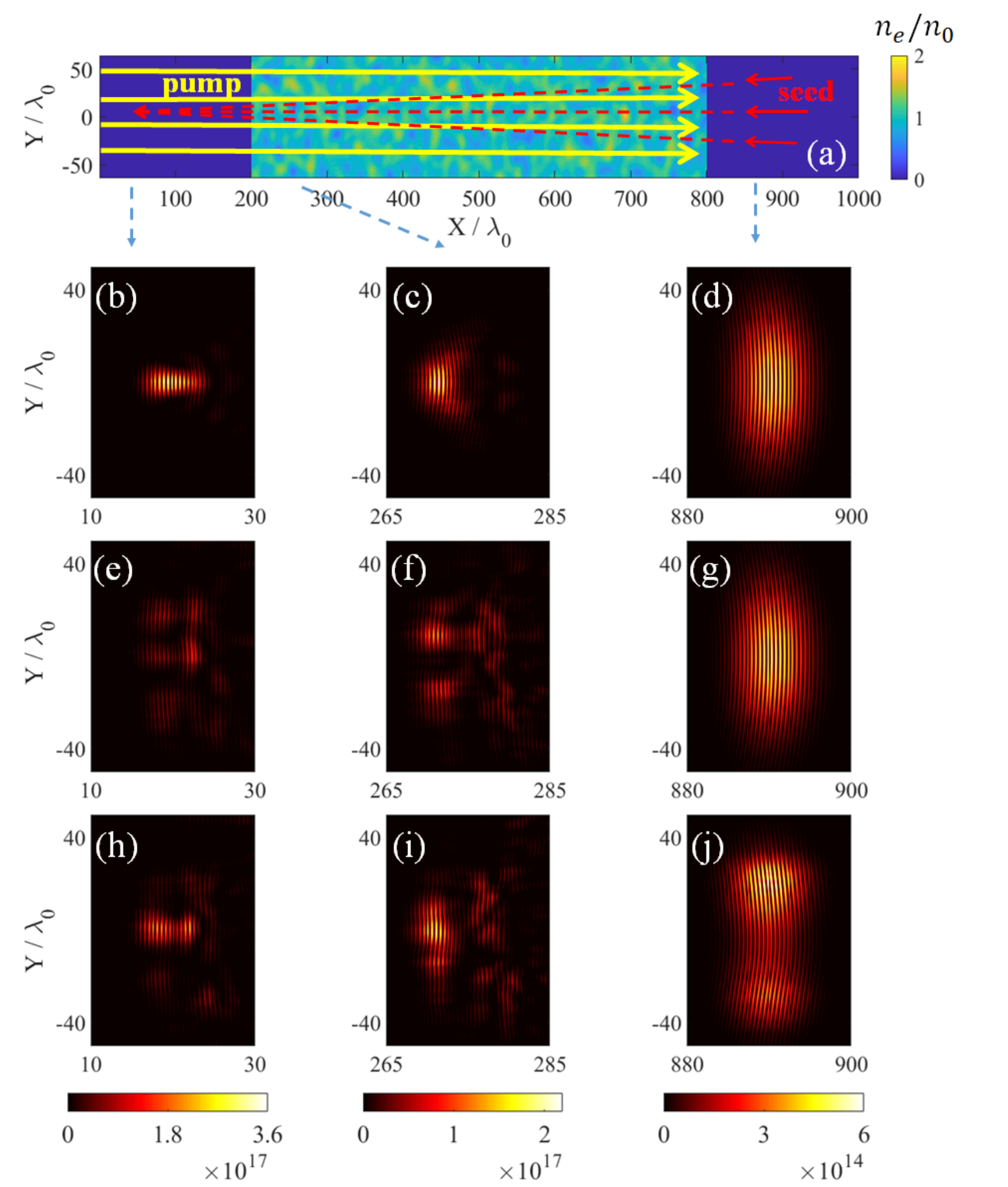}
\caption{Panel (a) shows the schematics of the PIC simulations, and the color mapping shows the inhomogeneous plasma density distribution. Panels (d)-(b) show three snapshots of the intensity profiles of a pre-focused seed pulse during BRA in homogeneous plasma; Panels (g)-(e) show a pre-focused seed pulse in inhomogeneous plasma; Panels (j)-(h) show a PC seed pulse in inhomogeneous plasma. The arrows below panel (a) illustrate the position of the snapshots.  }
\label{fig:intensity}
\end{figure}

The PIC simulations are conducted in two dimensions ($X$ and $Y$) using the full-relativistic kinetic code \texttt{EPOCH}~\cite{EPOCH2015}. For reference, we first demonstrate BRA of a tightly focused laser seed in a homogeneous plasma. As illustrated in Fig.~\ref{fig:intensity}(a), a pump pulse propagating in the x direction interacts with a counter-propagating seed pulse in a $0.45$~mm long plasma. The pump pulse has a wavelength $\lambda_\mathrm{ pump}=0.8 \mu$m and intensity $I_\mathrm{pump}=2.5\times10^{14}\rm W/cm^2$. The Gaussian-shape seed pulse has a wavelength $\lambda_{0} = 0.889 \mu$m and, after exiting the plasma, is focused at $X=0$ with a waist $w_0=8 \lambda_0$.  The electron number density of the homogeneous plasma is $n_0=0.01 n_c$, where $n_c=1.74\times10^{27}$m$^{-3}$ is the critical density for the pump pulse. The size of the simulation box is $1000\lambda_0\times128\lambda_0$ with 10 cells per $\lambda_0$ in both $X$ and $Y$ directions. Periodic boundary conditions is applied in the $Y$ direction and $64$ electrons per cell are placed between $X=200\lambda_0$ to $X=800\lambda_0$. The seed intensity profiles at different propagation distances are shown in Fig.~\ref{fig:intensity}(d), (c), and (b). 
The final pulse, as shown in Fig.~\ref{fig:intensity}(b), exhibits a smooth and regular tightly focused profile.

If the same pre-focused seed pulse is sent into an inhomogeneous plasma for amplification, the seed pulse becomes scattered and loses focusability after exiting the plasma~\cite{POP-Solodov2003, Palastro2015}. As an example, we introduce random plasma density perturbation with a long range correlation 
\begin{equation}
\frac{\left\langle\tilde{n}(\vec{r})\tilde{n}(\vec{r}+\vec{R})\right\rangle}{\left\langle\tilde{n}^2\right\rangle} = \exp{\left[-\frac{\pi (X^2+Y^2)}{l^2}\right]}, 
\label{eq:density}
\end{equation}
with $\tilde{n}/n_0=0.17$ and $l=40\lambda_0$. The simulation results show that the initially high-quality laser pulse [Fig.~\ref{fig:intensity}(g)] has completely lost its focusability at the exit [Fig.~\ref{fig:intensity}(f)]. The pulse profile at the focal plane [Fig.~\ref{fig:intensity}(e)] is separated into more than seven visible speckles and the peak intensity $1.3\times10^{17}\rm W/cm^2$ is only a fraction of the focused pulse shown in Fig.~\ref{fig:intensity}(b). 
The pulse energy in the central region reveals the reduced energy transfer efficiency.

We next replace the pre-focused seed pulse with a PC seed pulse and verify its performance of BRA in inhomogeneous plasma. The PC seed pulse, shown in Fig.~\ref{fig:intensity}(j), is obtained by sending a tightly focused pulse from $X=0$ through the inhomogeneous plasma and extracting the amplitude and phase information of the pulse when it reaches $X=900\lambda_0$. We then numerically take its PC and send it back to the plasma for BRA. Although the PC seed pulse exhibits a disrupted intensity profile at $X=900\lambda_0$, it gradually recovers its focusability when propagating through the same inhomogeneous plasma. When it exits the plasma at $X=200\lambda_0$, its intensity profile shows a regular focusing wavefront followed by several darker speckles, as shown in Fig.~\ref{fig:intensity}(i). The peak intensity of the main pulse is similar to that in Fig.\ref{fig:intensity}(c). The darker speckles which are separated from the leading pulse are the results of linear amplification, and hence they do not retain the PC wavefront. After propagating in vacuum, the amplified pulse at $X=0$ is able to focus into the short pulse [Fig.~\ref{fig:intensity}(h)] together with a dark halo. The peak intensity reaches $2.5\times10^{17}\rm W/cm^2$, which is almost twice larger than the amplified pulse using a pre-focused seed in the same inhomogeneous plasma. 

\begin{figure}[htbp]
	\centering
	\includegraphics[width=\linewidth]{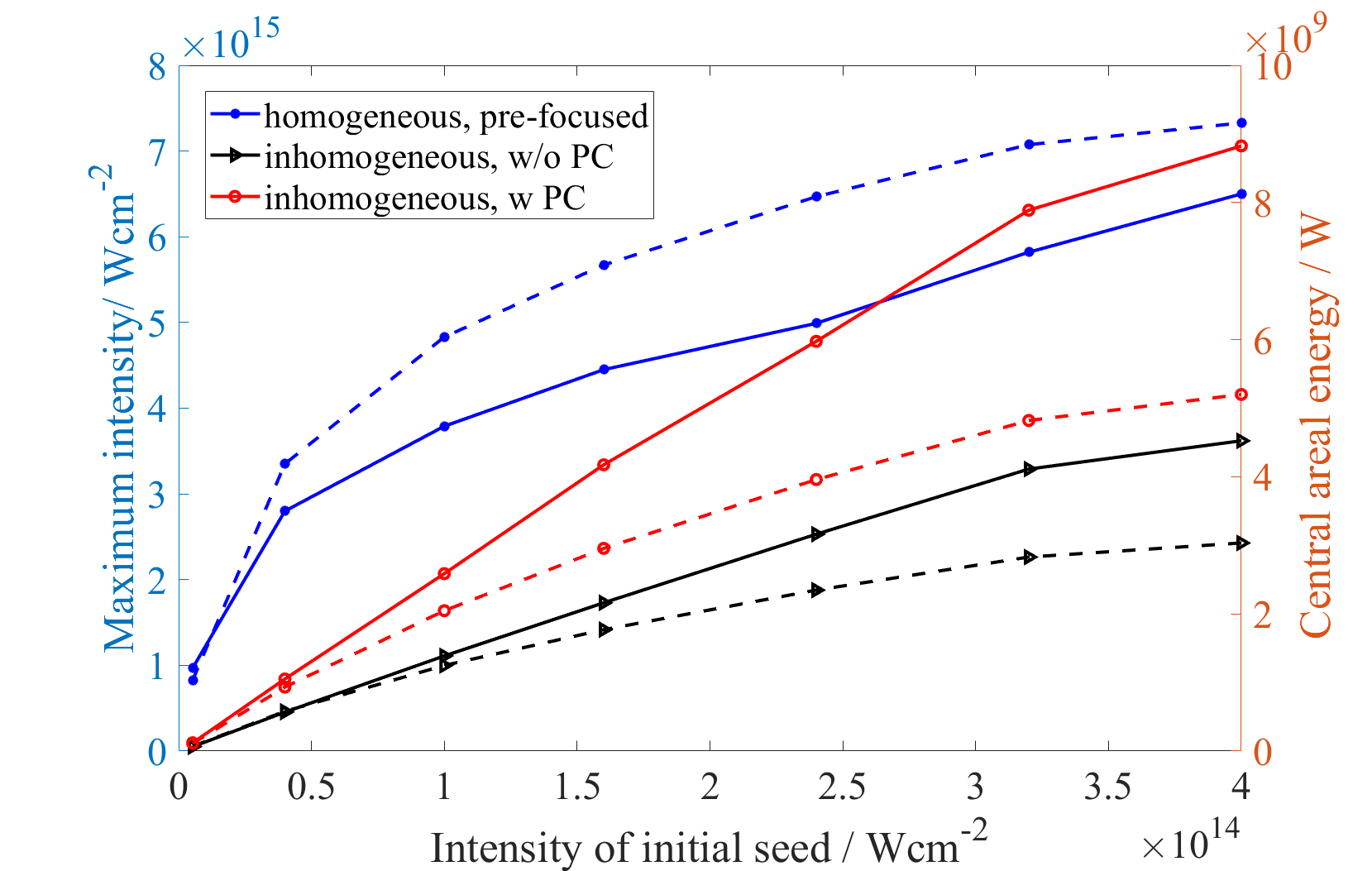}
	\caption{The peak intensity (left axis, solid curves) and pulse energy (right axis, dashed curves) of the amplified seed at the focal plane with varying initial seed intensities. The three groups of curves are the results of using pre-focused seed in homogeneous plasma (blue dots), pre-focused seed in  inhomogeneous plasma (black triangles) and PC seed in inhomogeneous plasma (red circles), respectively. }
	\label{fig:nonlinear}
\end{figure}

Further increasing the energy transfer efficiency and eliminating the unfocused speckles could be achieved by reducing the length of linear amplification stage, \eg with stronger seed pulses. A criteria for reaching the advanced nonlinear stage is that the seed intensity exceeds the pump intensity. For verification, we change the initial seed intensity from below the pump intensity ($2.5\times10^{14}\rm W/cm^2$) to several times above that and repeat the PIC simulations sketched in Fig.~\ref{fig:intensity}. The results shown in Fig.~\ref{fig:nonlinear} compare the peak intensity and the pulse energy of the final focused amplified pulse. At very low seed intensity, BRA works mostly in the linear regime and the amplified pulse does not retain the wavefront of the seed pulse. As expected, the PC seed pulse performs similarly to a pre-focused pulse. With higher initial seed intensities, the 
length of linear amplification decreases and the length of nonlinear amplification increases. From Fig.~\ref{fig:nonlinear}, we find that the amplified pulse intensity of a PC seed grows quicker than that of a pre-focused pulse. It approaches the pulse intensity from a homogeneous plasma when $I_\mathrm{seed} \sim I_\mathrm{pump}$. More interestingly, the simulation results in the region $I_\mathrm{seed} > I_\mathrm{pump}$ demonstrate that a PC seed pulse could be amplified to a pulse with higher intensity than a pre-focused pulse despite of the plasma inhomogeneity. 
We also compare the amplified pulse energy in the $(8\lambda_0)^2$ region for different BRA schemes. The results in Fig.~\ref{fig:nonlinear} yield the same conclusion that PC seed pulses can gain more energy than pre-focused pulses in the nonlinear amplification regime, although they cannot restore as much energy as the ideal case of homogeneous plasma.

The 2D PIC simulations clearly demonstrate the capability of retaining focusability of a strong PC seed pulse in BRA. 
Since the wavefront of the amplified pulse resembles that of the initial seed pulse, the pulse can even be focused behind an aberrating medium. 
For this purpose, the seed pulse is to be created by PC of a reference beam---the radiation of the target after passing through the aberrating media and the plasma. 
The PC seed is then amplified by a pump pulse through BRA while retaining its focusability and deposits its energy to the target region. 
This scheme may have advantages in applications like laser fusion and high-power radiation transmission where optical aberrations due to plasma/air density fluctuation impedes the focusing of laser power.

The full potential of this method can be realized only by first creating strong PC seed pulses. 
The time for seed creation and round-trip propagation cannot exceed the time scale of the aberration dynamics. 
For plasmas, the relevant time scale is determined by ion motion, which is in the $\mathrm{ns}$ scale for $\mathrm{mm}$-size plasmas. 
The methods with stimulated Brillouin scattering~\cite{1972JETPL15109Z} and degenerated four-wave mixing~\cite{1977Hellwarth} in crystal are too slow and cannot resist high pulse intensities. 
A potentially viable method for PC is using stimulated emission~\cite{He_OL97,He_PRA2008,2007preLee,2012olDing}: an amplifying medium with population inversion is placed in a scattering medium; the PC backscattered light is amplified when propagating. The lasing process contributes both to increasing the output pulse energy and also to reducing the formation time of PC. The reported experiments have demonstrated PC of a $160$~fs pulse in the $\mu\mathrm{J}$ energy level. An upgrade of such a setup might reach both the required time-scale and pulse intensity.

This work was supported by NNSA Grants No. DE-NA0002948, No. DE-NA0003871, and AFOSR Grant No. FA9550-15-1-0391.

\bibliography{ref}



\end{document}